\documentstyle[hip-artc]{article}
\volnumber{11}  \edyear{2000}  \frompage{35} \topage{41}                 
\recrevdate{7 September 1999}                                            
\def\beq{\begin{equation}}
\def\eeq{\end{equation}}

\def\a{\alpha}
\def\ba{\beta}
\def\g{\gamma}
\def\d{\delta}
\def\l{\lambda}
\def\n{\nabla}
\def\e{\epsilon}
\def\s{\sigma}
\def\prt{\partial}

\def\rightmark{Coset-Space String Compactification}
\def\leftmark{A.M.~Gavrilik}

\title{Coset-Space String Compactification Leading \\
       to 14 Subcritical Dimensions}
\authors{
{\twerm A.M.~Gavrilik$^a$
}\\[2.812mm]
{\normalsize
\hspace*{-8pt} Bogolyubov Institute for Theoretical Physics \\
 03143 Kiev-143, Ukraine\\[0.2ex]
}}

\abstract{
Using sigma-model approach, we study a class of coset spaces
with torsion which compactify the $D=26$ closed bose-string
theory. Requiring also that massless chiral fermions
arise from the geometry/topology of coset space, we are left with
the unique possibility:
it implies $D=14$ subcritical dimensions and 
the isometry group $G_2\times G_2$.
}

\begin{document}

\maketitle
\setcounter{page}{35}

\vspace{-3mm}

\section{Introduction}

The common belief now is that an adequate construct of TOE
(theory of everything) will employ some version of string
theory \cite{GSW}. With impact on 3+1 phenomenology, various
compactifications were studied \cite{GSW,Ke} in the context of
ten dimensional (super) strings. These latter however
are known to result \cite{Fr}, via dimensional reduction
using $16$-torus, from the $D=26$ bosonic string theory which is
thus the only one preserving the
{\it status of unique string theory}.
No doubt, it is of great importance to study other (than toroidal)
ways of bose-string compactification starting directly from $D=26$.
For this purpose, more than a decade ago in \cite{Ga} we proposed
to use certain compact coset spaces $K$ (of dimension $d$) with
nontrivial torsion for getting compactifications of the form
\beq
M^{26}\to M^{26-d}\times K^d ,
\eeq
and studied their properties using sigma-model approach
(previously applied for superstring compactifications, see
e.g., \cite{GSW,Ke,CL}).
Here, after a sketch of main facts from \cite{Ga}, we present
arguments in favour of definite unique compactification which
fixes (i) the number $14$ of subcritical dimensions (SCDs),
(ii) gauge group in $D=14$ stemming from isometry of
compactifying coset space, and provides chiral fermions.

\newpage
\section{Sigma-Model $\beta$-Functions and Consistency Equations}

Action for string propagation in the background
metric, antisymmetric-tensor and dilaton
fields $G_{\mu\nu}$, $B_{\mu\nu}$ and $\Phi$ is taken as \cite{FT-CFMP}
\vspace{2mm}
\[
\hspace{-78mm}  S=S_G+S_B+S_{\Phi}
\vspace{-1mm}
\]
\beq
=\frac{1}{4\pi{\alpha}'}  \int d^2 \xi \left[
 G_{\mu\nu}(X)\partial_\a X^\mu \partial_\a X^\nu
+{\e }^{\a \ba } B_{\mu\nu }(X)
                {\prt }_{\a } X^{\mu } {\prt }_{\ba } X^{\nu }
\right]
+ S_{\Phi}.                           \label{1}
\eeq
Conformal (Weyl) anomaly of such generalized 2d $\s$-model
consists of three terms,
\beq
C.A.={\ba}^G_{\mu\nu} {\prt}_{\a }X^{\mu} {\prt}_{\a} X^{\nu} +
{\ba}^B_{\mu\nu} {\e }^{\a {\ba} } {\prt}_{\a} X^{\mu} {\prt}_{\ba} X^{\nu}
+{\ba}^{\Phi}\frac14 \sqrt{g} R^{(2)}.
\label{2}
\eeq
Clearly, it should vanish for the consistency of string propagation in
these backgrounds, that implies vanishing of their beta-functions:
${\ba}_{\mu\nu}^G = {\ba}_{\mu\nu}^B = {\ba}^{\Phi} = 0.$
Remark that the phenomenon of "geometrostasis" due to presence
of torsion \cite{CZ} is crucial in order to guarantee
such vanishing (or, the appearance of nontrivial fixed point in
the space of $G_{\mu\nu},\ B_{\mu\nu },\ {\Phi},$
treated as "couplings"). These consistency (the vanishing) relations
are also interpreted as the equations for string propagation in such
external fields. To describe a class of consistent coset-space
string compactifications, we first exploit these equations
to lowest order, namely \cite{FT-CFMP}
\vspace{-1.0mm}
\[
\hspace{-33mm}
 {\ba}^G_{\mu\nu} 
= R_{\mu\nu} -
       \frac14 H_{\mu}^{\ {}\l\s}H_{\nu\l\s}
         + 2 \n_{\mu}\n_{\nu}\Phi \\
\]
\vspace{-3.6mm}
\beq
\hspace{-48mm}
{\ba}^B_{\mu\nu} 
=\n_{\l} H^{\l}_{\ {}\mu\nu}
         -2 \n_{\l}\Phi H^{\l}_{\ {}\mu\nu} \\
\eeq
\vspace{-3.0mm}
\[
\hspace{1mm}
{\ba}^{\Phi} 
=\frac{1}{\a'}\frac{D-26}{48\pi^2}+
\frac{1}{16\pi^2}\left\{- R +
\frac{1}{12}H^2 + 4 (\n\Phi)^2 - 4\n^2\Phi\right\}
\]
\vspace{-0.6mm}
where $R_{\mu\nu}$ and $R$ are Ricci tensor and scalar curvature of
the $\s$-model target space,
$H_{\mu\nu\l}\equiv 3\partial_{[\mu}B_{\nu\l ]}$
and $H^2 \equiv H_{\a\ba\gamma} H^{\a\ba\gamma}$
(the 3-form $H$ is linked to torsion $T$).


In all the treatment below we set $\Phi=const.$
The presence of torsion modifies both the connection and Riemann tensor
according to
\beq
\begin{array}{cll}
\Gamma^{\mu (sym.)}_{\ \nu\rho} & \longrightarrow &
     {\tilde{\Gamma}}^{\mu}_{\ \nu\rho}
\equiv\Gamma^{\mu (sym.)}_{\ \nu\rho} - H^{\mu}_{\ \nu\rho} , \\
R_{\mu\nu\rho\s}            & \longrightarrow &
{\tilde R}_{\mu\nu\rho\s}
= R_{\mu\nu\rho\s} + \n_{\rho}H_{\mu\nu\s} - \n_{\s}H_{\mu\nu\rho}
+ H_{\l\mu\rho}H^{\l}_{\ {}\s\nu} - H_{\l\mu\s}H^{\l}_{\ {}\rho\nu} , \\
  R_{\mu\nu}  &   \longrightarrow  & {\tilde R}_{\mu\nu} =
{\tilde R}_{[\mu\nu ]} + {\tilde R}_{(\mu\nu )} ,
\end{array}
\eeq                             \label{5}
and the requirement of absence of conformal anomaly to one-loop
implies vanishing of the symmetric and antisymmetric parts
of generalized Ricci tensor \cite{CZ,CL}:
\beq
{\ba}^G_{\mu\nu} = {\tilde R}_{(\mu\nu )} = 0 ,\ \ \ \ \ \
{\ba}^B_{\mu\nu} = {\tilde R}_{[\mu\nu ]} = 0 .
\eeq
Thus, at this order, one searches for the desired solutions among
$\widetilde{\rm Ricci}$-flat
spaces, to be extracted from the variety of coset spaces.
Also, we have to ensure ${\ba}^{\Phi} = 0.$

\section{The Relevant Coset Spaces}

The  vielbeins $e^a(y)=e^a_{\mu}(y) dy^{\mu}$ for a coset-space
$G/H$ are obtained from the Lie-algebra valued 1-form
$V(y)\equiv L^{-1}(y) d L(y) = e^a(y) I_a + e^i(y) I_i$
in which the generators $I_i,\ i=1,2,...,{\rm dim}\ H$
(or $I_a,\ a=1,2,...,{\rm dim}\ G/H$) generate Lie algebra
of the subgroup $H$ (or tangent space of $G/H$ at the unit
of Lie group $G$) and satisfy
\beq
[I_a , I_b ]  =  {f^i }_{ ab} I_i + {f^c }_{ ab} I_c ,\ \ \ \ \ \
[I_i , I_j ]  =  {f^k }_{ ij} I_k ,  \ \ \ \ \ \ 
[I_i , I_a ]  =  {f^b }_{ ia} I_b .     
\eeq
The following main classes of coset spaces are to be mentioned.

(i) Group manifolds (with a Lie group $G$)
equivalent to ${(G\times G)}/{G}$;

(ii) Symmetric spaces [typical representatives (TRs): $N$-spheres
$S^N=\frac{SO(N+1)}{SO(N)}$

\hspace{4mm} and Grassmannians
$Gr_{N,k}=\frac{SO(N)}{SO(N-k)\times SO(k)}$].
For these, ${f^c }_{ ab}=0$ in (7);

(iii) Nonsymmetric spaces, splitting into

\ \ \ \ \ \ $\bullet$ isotropy-reducible ones [TRs are Stiefel manifolds
                         ${SO(N)}/{SO(N-k)}$],

\ \ \ \ $\bullet\hspace{-0.2mm}\bullet$ isotropy irreducible ones
           [TR is the Berger manifold ${Sp(2)}/{SU(2)}$].

\noindent
Symmetric spaces (ii) possess no torsion; group manifolds
admit torsion, but are inappropriate since they
do not yield \cite{WuZ} chiral fermions. Thus,
we are left with the class (iii). But, the {$\bullet $}-subclass
contains coset manifolds with complicated structure of their spaces
of Riemannian metrics described by many parameters (``moduli'').
We restrict ourselves to the case of minimal possible number
$n=1$ of moduli.
Note that moduli space of both symmetric spaces and nonsymmetric
isotropy-irreducible (NSII) ones, is 1-dimensional (homothety of
metrics is the relevant parameter). Thus, we choose the
{$\bullet\bullet $}-subclass of NSII coset spaces.

These coset spaces were completely described
by Manturov and Wolf \cite{Ma,Wo}.

\vspace{5mm}
{\small
\begin{center}
{\bf Table 1.
NSII
(Manturov-Wolf)
coset spaces of dimension $\le 24$}
\end{center}

\vspace{-6mm}
\begin{center}
\begin{tabular}{llll}
\hline
Coset space &  Reduction  & Coset space  & Reduction \\
$G/H$ &  $M^{26}\to$ &  $G/H$  & $M^{26}\to$\\

\hline
$K^6=G_2/SU(3) $ & $\to M^{20}\times K^6$  &
  $K^{15}_A=Sp(3)/Sp(1)\times SO(3)$  &  $\to M^{11}\times K^{15}_A$\\
$K^7_A=Sp(2)/SU(2) $ & $\to M^{19}\times K^7_A$  &
  $K^{15}_B=SO(8)/Sp(1)\times Sp(2)$  &  $\to M^{11}\times K^{15}_B$\\
$K^7_B=SO(7)/G_2 $ & $\to M^{19}\times K^7_B$  &
   $K^{20}=SO(8)/SU(3)$  &  $\to M^{6}\times K^{20}$ \\
$K^{11}=G_2/SO(3) $ & $\to M^{15}\times K^{11}$  &
   $K^{24}=SU(6)/SU(2)\times SU(3)$  &  $\to M^{2}\times K^{24}$\\
\hline
\end{tabular}
\end{center}
}

\vspace{5mm}
Setting ${T^c }_{ ab}=\eta{f^c }_{ ab}$ in the Maurer-Cartan equations
for (generalized) curvature and torsion 2-forms, one gets expressions
for $\tilde{R}^a_{\ bcd}$ and $\tilde{R}_{ab}$ in terms of structure
constants and the coefficient $\eta$. Using the resulting
formulas \cite{CL}
\vspace{-2mm}
\beq
\tilde{R}^a_{\ bcd}= {f^a }_{ bi}{f^i }_{ cd}
  + \frac12(1+\eta){f^a }_{ be} {f^e }_{ cd}
+ \frac14 {(1+\eta)^2} [ {f^a }_{ ce} {f^e }_{ db}
- {f^a }_{ de} {f^e }_{ cb}] ,
\eeq
\vspace{-4mm}
\beq    \hspace{-18mm}
\tilde{R}_{ab}= {R}_{ab} - \frac14 {T^c }_{ ad} {T^d }_{ cb}
 = {f^e }_{ ai} {f^i }_{ eb}
+ \frac14 (1-\eta)^2 {f^c }_{ ad} {f^d }_{ cb},
\eeq
we calculate the metric $\ba$-function
for the case \cite{Ga2} of Berger manifold (BM)
to obtain the 1-loop result \cite{Ga}
${\ba}^{G (1)}_{ab}=(13.5-\frac32\eta^2)\d_{ab}$. So, the choice
$\eta=\pm 3$ ensures vanishing of the metric $\ba$-function.
Also, one can argue for the BM that ${\ba}^B_{ab}=0$
(in fact, this holds for all coset spaces in table 1).
As shown in \cite{Ga}, for the BM 
there exists such a connection (such value of $\eta$, i.e.,
$\eta =\pm 3\sqrt{3}$) that ensures 
${\ba}^{\Phi} = 0$.
Similar 1-loop analysis can be done \cite{Ga} for all other
NSII coset spaces.

As shows table 1, the number 10 of SCDs
as well as the observed number 4, are forbidden. This
dictates the necessity of (at least) two separate stages
of compactification, the second one implying presence
of certain nonabelian gauge field generated by isometry of
the first-stage coset space. However, there is an alternative
- to consider not only simple, but also product-type compactifications,
see Section 5, of the form $K^i \times K^{j}$,\
$K^i \times K^{j}\times K^l$,\ and
$K^i \times K^{j}\times K^l\times K^m$, with factors taken
from the table 1.

\section{Extension to the Two-Loop Order}

There was an extensive study of $\beta$-functions of these
generalized $\s$-models to two-loop order, see \cite{MT} and
references therein. While there was no renormscheme (RS) ambiguity,
to this order, with absent $B$-field, essential RS-ambiguities
do appear when $B_{\mu\nu}$ is included. The two-loop expressions for
${\beta}^G_{\mu\nu}$,\ ${\beta}^B_{\mu\nu}$ and ${\beta}^\Phi$,\
obtained in \cite{MT}, depend on the 3 extra parameters $f_1, p_1, p_2$
characterizing RS-dependence.
Let us quote the expression for ${\beta}^{G (2)}_{\mu\nu}$
in a particular RS:
\beq
{\beta}^G_{\mu\nu}[f_1\!\!=\!\!-1] = {\a}'{\tilde R}_{(\mu\nu)} +
\frac{{\a'}^2}{2}
\Bigl\{
{\tilde R}^{\a\beta\g} {}_{(\nu}{\tilde R}_{\mu )\a\beta\g}
- \frac12
{\tilde R}^{\beta\g\a} {}_{(\nu}{\tilde R}_{\mu )\a\beta\g} +
\frac12 {\tilde R}_{\a(\mu\nu )\beta} (H^2)^{\a\beta}
\Bigr\}.
\eeq
Straightforward calculation, using (10), for the concrete case
of BM $K^7_A$ shows:  With the canonical ($\eta=-1$) connection,
this coset space
cannot provide a solution (i.e., no RS exists) of the equation
${\beta}^{G (2)}_{\mu\nu} = 0$. On the contrary, at certain
connections with $\eta\ne -1$, even in the scheme $f_1=-1$
there are solutions (that follows from calculations
involving (10)). Like in \cite{CL}, set $\l = \alpha'/r^2$.
Besides the trivial solution $\l = 0,\ \eta=\pm 3$, there exists
the one for which $\l$ is nonzero and $1> \l > 0$:
\beq
\eta=\pm 3 - \d, \ \ \ \ \ \  \l =
{P_2(\eta )}/{P_4(\eta)}
\eeq
where $\d$ is a small parameter; $P_2(\eta )$ and $P_4(\eta )$ are
concrete polynomials of 2nd and 4th order in $\eta$.

From (11) it follows that the 2-loop consistency condition
${\beta}^{G (2)}_{\mu\nu} = 0$\
imposes concrete relation between the magnitude
of torsion and the size of manifold (compactification radius).
In general, solutions of ${\beta}^G_{\mu\nu}={\beta}^\Phi=0$
for a NSII coset space are found by solving the system of two
equations $f(G/H; \l,\eta)=0$,\ $h(G/H; \l,\eta)=0$, with
definite scalar functions dependent on $G/H$.

\section{Product-Type Compactifications}

Similar analysis can be applied to the case of product spaces.
The following table lists string compactifications
on product coset-spaces.

\vspace{4mm}
{\small
\begin{center}
{\bf Table 2. List of product compactifications}
\end{center}
}

\begin{center}
\begin{tabular}{cccl}
\hline
       {} &     {}     &  Subcritical  &  Euler  \\
$d$      &  $[G/H]^d$ &  dimension:   & charact- \\
       {} &     {}     &  dim$=26-d$ &  eristics \\
\hline
$1,2,3,4,5$ &  {\bf None}  & 25,24,23,22,21  & {}  \\
\hline
       $6$  &  $K^{6}$ &  $20$    &  $\chi\ne 0$   \\
\hline
       $7$  &  $K^{7}_A ,\ \ \ \ \ \ K^{7}_B$ &  $19$  &  {} \\
\hline
$8,9,10$ & {\bf None}  & 18,17,16 & {} \\
\hline
       $11$  &  $K^{11}$ &  $15$ & {} \\
\hline
       $12$  &  $K^{6}\times K^6$         &  ${\bf 14}$  &  $\chi\ne 0$\\
\hline
$13$  &  $K^6\times K^{7}_A , \ \ \ \ K^6\times K^{7}_B$ &  $13$  &  {} \\
\hline
$14$  &
$K^7_A\times K^{7}_A ,\ \ \ \ K^7_A\times K^7_B , \ \ \ \
       K^7_B\times K^7_B$ & $12$ & {} \\
\hline
       $15$  &  $K^{15}_A, \ \ \ \ \ \ K^{15}_B$ &  $11$   &   {}  \\
\hline
       $16$  &     {\bf None }   &  ${\bf 10}$   &   {}  \\
\hline
       $17$  &  $K^6 \times K^{11}$ &  $9$   &   {}   \\
\hline
       $18$  &  $K^6\times K^6\times K^6 $ &  $8$   &   $\chi\ne 0$   \\
          {}  &  $K^{7}_A\times K^{11} ,\ \ \ \  K^{7}_B\times K^{11}$
                                                   &  {}   & $\chi = 0$ \\
\hline
       $19$  &  $K^6\times K^6\times K^{7}_{A,B}$ &  $7$   &   {}   \\
\hline
       $20$  &  $K^{20} ,\ \ \ \ K^6\times (K^{7}\times K^7)_{AA,AB,BB}$
                                          &  $6$   &   {}   \\
\hline
       $21$  &  $K^6\times K^{15}_{A,B} ,$    &  $5$   &   {}   \\
       {}  &  $(K^7\times K^7\times K^{7})_{AAA,AAB,ABB,BBB}$ & {} & {} \\
\hline
       $22$  &  $K^{11}\times K^{11} ,\ \ \ \
(K^7\times K^{15})_{AA,AB,BA,BB}$ &  $ {\bf 4}$  &  ${\bf \chi = 0}$  \\
\hline
       $23$  &  $K^{6}\times K^{6}\times K^{11}$ &  $3$   &   {}   \\
\hline
  $24$  &  $K^{24} , \ \ \ \ K^6\times K^6\times K^6\times K^6 $ &
                                                       $2$ & $\chi\ne 0$ \\
  {} &  $K^6\times K^{11}\times K^7_{A,B} $ & {} & $\chi = 0$ \\
\hline
       $25$  &  $K^6\times K^6\times K^6\times K^{7}_{A,B} ,$ &  $1$ & {}\\
       {}  &  $K^{11}\times (K^7\times K^{7})_{AA,AB,BB}$ & {} & {}\\
\hline
$26$ & $K^6\times K^{20} ,\ \ \ \ (K^{11}\times K^{15})_{A,B}$ & $0$ & {}\\
     & $K^6\times K^6\times (K^{7}\times K^{7})_{AA,AB,BB} ,$ & {}  & {} \\
\hline
\end{tabular}
\end{center}

\newpage

Again, there is no compactification leading to ten SCDs.
Although the case of $\ 4\ $SCDs is admissible ($d = 22$ line in the table),
the five compactifying coset spaces do not yield chiral fermions:
as can be shown, the Euler characteristics
$\chi = 0$ for all these $d = 22$ coset spaces.
Thus, we confirm: there {\it must be (at least) two stages}
of coset-space compactification of the considered type, to obtain
the realistic 4 dimensions with necessary chiral fermion
(quark-lepton) families.
The unique candidate for such first stage of coset-space
string compactification, as shows the table, is nothing but
the $d = 12$ space $K^6 \times K^6$ with $\chi = 4$,
which leads to 14 dimensions and chiral fermions therein.

\section{Concluding Remarks}

1) We arrived at the necessity, even using product coset spaces,
of at least 2 stages of compactification. The number 10 of SCDs
is forbidden, from which we conclude that the considered
compactifications supply string vacua differing from the known
$10 d$ (super)string theories. 2) In fact,
$H\ne T$ since ${\rm d}T\ne 0$, which follows from explicit calculation.
This dictates to include \cite{CL} the Lorentz Chern-Simons 3-form
$\omega_{3L}$. 3) Dilaton field $\Phi$, playing a special role,
requires detailed account of string loops, see \cite{CLNY}.
4) Besides the approximate approach used so far,
nonperturbative construction of a complete solution
is certainly needed. A question also arises whether relevant
theory in 14 dimensions, based on this unique compactification and
yet to be constructed, may have some connection to the
(bosonic sector of) recently proposed \cite{Ba} supersymetric
theories in 14 (i.e. 11+3) dimensions.

\section*{Acknowledgements}
The author thanks Prof. J.Klauder and Prof. M.Tonin for their interest
in the topic. He acknowledges Prof. I.Lovas and Prof. T.Hadhazy for
their kind hospitality during the conference. This work was supported
in part by the CRDF award No. UP1-309 and by the grant INTAS-93-1038-ext.

\section*{Note}
\begin{notes}
\item[a]
E-mail: omgavr@bitp.kiev.ua
\end{notes}

\vfill\eject

\begin{thebibliography}{99}

\bibitem{GSW} M.B.~Green, J.H.~Schwartz and E.~Witten,
{\it Superstring Theory}, Cambridge Univ.Press, Cambridge, 1987, p. 1065.

\bibitem{Ke} S.~Ketov, {\it Introduction to the Quantum Theory
of Strings and Superstrings}, Nauka, Novosibirsk, 1990, p. 369,
in Russian.

\bibitem{Fr}P.G.O.~Freund,
{\it Phys. Lett.} {\bf 151B} (1985) 387;
A.~Casher, P.~Englert, H.~Nicolai and A.~Taormina,
{\it Phys. Lett.} {\bf 162B} (1985) 121.

\bibitem{Ga} A.M.~Gavrilik, Preprint ITF--87--82P, Kiev, 1987.

\bibitem{CL} L.~Castellani and D.~L\"ust,
{\it Nucl. Phys.} {\bf B296} (1988) 143.

\bibitem{FT-CFMP} E.S.~Fradkin and A.A.~Tseytlin,
{\it Phys. Lett.} {\bf 158B} (1985) 316; {\bf 160B} (1985) 69;
C.~Callan, D.~Friedan, E.~Martinec and M.~Perry,
{\it Nucl. Phys.} {\bf B262} (1985) 593.

\bibitem{CZ}
T.L.~Curtright and C.~Zachos,
{\it Phys. Rev. Lett.} {\bf 53} (1984) 1799;
E.~Braaten, T.L.~Curtright and C.~Zachos
{\it Nucl. Phys.} {\bf B260} (1985) 630.

\bibitem{WuZ} Y.-S.~Wu and A.~Zee,
{\it J. Math. Phys.} {\bf 25} (1984) 2696.

\bibitem{Ma} O.V.~Manturov,
{\it DAN SSSR} {\bf 141} (1961) 792; ibid. 1034.

\bibitem{Wo} J.A.~Wolf,
{\it Acta Math.} {\bf 120} (1968) 59.

\bibitem{Ga2} A.M.~Gavrilik,
{\it Teor. Mat. Fiz.} {\bf 65} (1985) 155.

\bibitem{MT} R.R.~Metsaev and A.A.~Tseytlin,
{\it Nucl. Phys.} {\bf B293} (1987) 385.

\bibitem{CLNY} C.G.~Callan, C.~Lovelace, C.~Nappi and S.~Yost,
{\it Nucl. Phys.} {\bf B288} (1987) 525.

\bibitem{Ba} I.~Bars,
{\it Phys. Lett.} {\bf 403B} (1997) 257;
E.~Sezgin,
{\it Phys. Lett.} {\bf 403B} (1997) 265;
 M.~Nishino, {\em Nucl. Phys.} {\bf B523} (1998) 450;
 I.~Rudychev, E.~Sezgin and P.~Sundell, 
         {\em Nucl. Phys. Proc. Suppl.} {\bf 68} (1998) 285.

\end{thebibliography}
\end{document}